\documentclass[aps,prb,twocolumn,superscriptaddress,groupedaddress,showpacs]{revtex4}
\usepackage{amsmath,amsfonts,amssymb,bm}
\usepackage{graphicx}
\usepackage{dcolumn}
\usepackage{epsfig}
\usepackage{color,xcolor}
\usepackage{units}
\usepackage[amssymb]{SIunits}

\begin{document}
\title{Perovskite-type cobalt oxide at the multiferroic Co/Pb Zr$_{0.2}$Ti$_{0.8}$O$_{3}$ interface}
\author{K. Mohseni}
\affiliation{Max-Planck-Institut  f\"ur Mikrostukturphysik, Weinberg 2, 06120 Halle, Germany}
\author{A. Polyakov}
\affiliation{Max-Planck-Institut f\"ur Mikrostukturphysik, Weinberg 2, 06120 Halle, Germany}
\author{H.\,B. Vasili}
\affiliation{Alba Synchrotron Light Source, 08290 Cerdanyola del Vall{\'e}s Barcelona, Spain}
\author{I.\,V. Maznichenko}
\affiliation{Institute of Physics, Martin Luther University Halle-Wittenberg, 06099 Halle, Germany}
\author{S. Ostanin}
\affiliation{Institute of Physics, Martin Luther University Halle-Wittenberg, 06099 Halle, Germany}
\author{A. Quindeau}
\affiliation{Max-Planck-Institut  f\"ur Mikrostukturphysik, Weinberg 2, 06120 Halle, Germany}
\author{N. Jedrecy}
\affiliation{Institut des Nano Sciences de Paris (INSP), Sorbonne Universite, CNRS UMR 7588, 4 Place Jussieu, 75252 Paris Cedex 05, France}
\author{E. Fonda}
\affiliation{SOLEIL, L'Orme des Merisiers, 91190 Saint-Aubin, France}
\author{L.\,V. Bekenov}
\affiliation {Institute of Metal Physics, Vernadsky Street, 03142 Kiev, Ukraine}
\author{V.\,N. Antonov}
\affiliation {Institute of Metal Physics, Vernadsky Street, 03142 Kiev, Ukraine}
\author{P. Gargani}
\affiliation{Alba Synchrotron Light Source, 08290 Cerdanyola del Vall{\'e}s Barcelona, Spain}
\author{M. Valvidares}
\affiliation{Alba Synchrotron Light Source, 08290 Cerdanyola del Vall{\'e}s Barcelona, Spain}
\author{I. Mertig}
\affiliation{Institute of Physics, Martin Luther University Halle-Wittenberg, 06099 Halle, Germany}
\author{S.\,S.\,P. Parkin}
\affiliation{Max-Planck-Institut  f\"ur Mikrostukturphysik, Weinberg 2, 06120 Halle, Germany}
\author{A. Ernst}
\affiliation{Max-Planck-Institut  f\"ur Mikrostukturphysik, Weinberg 2, 06120 Halle, Germany}
\affiliation{Institut f\"ur Theoretische Physik, Johannes Kepler Universit\"at, A 4040 Linz, Austria}
\author{H.\,L. Meyerheim}\email{holger.meyerheim@mpi-halle.mpg.de}
\affiliation{Max-Planck-Institut  f\"ur Mikrostukturphysik, Weinberg 2, 06120 Halle, Germany}

\date{\today}
\begin{abstract}
  Magnetic Tunnel Junctions whose basic element consists of
  two ferromagnetic electrodes separated by an insulating non-magnetic
  barrier have become intensely studied and used in non-volatile
  spintronic devices. Since ballistic tunnel of spin-polarized
  electrons sensitively depends on the chemical composition and the
  atomic geometry of the lead/barrier interfaces their proper design
  is a key issue for achieving the required functionality of the
  devices such as e.g. a high tunnel magneto resistance. An
  important leap in the development of novel spintronic devices is to
  replace the insulating barrier by a ferroelectric which adds new
  additional functionality induced by the polarization direction in
  the barrier giving rise to the tunnel electro resistance (TER).  The
  multiferroic tunnel junction
  Co/PbZr$_{0.2}$Ti$_{0.8}$O$_{3}$/La$_{2/3}$Sr$_{1/3}$MnO$_3$
  (Co/PZT/LSMO) represents an archetype system for which - despite
  intense studies - no consensus exists for the interface geometry and
  their effect on transport properties. Here we provide the first
  analysis of the Co/PZT interface at the atomic scale using
  complementary techniques, namely x-ray diffraction and extended
  x-ray absorption fine structure in combination with x-ray magnetic
  circular dichroism and ab-initio calculations. The Co/PZT interface
  consists of one perovskite-type cobalt oxide unit cell
  [CoO$_{2}$/CoO/Ti(Zr)O$_{2}$] on which a locally ordered cobalt film
  grows. Magnetic moments (m) of cobalt lie in the range between m=2.3 and
  m=2.7$\mu_{B}$, while for the interfacial titanium atoms they are
  small (m=+0.005 $\mu_{B}$) and parallel to cobalt which is
  attributed to the presence of the cobalt-oxide interface layers.
  These insights into the atomistic relation between interface and
  magnetic properties is expected to pave the way for future high TER
  devices.
\end{abstract}
\keywords{Multiferroic Interface, Co/Pb Zr$_{0.2}$Ti$_{0.8}$O$_{3}$,
  Surface X-Ray Diffraction, X-Ray Absorption Fine Structure,
  X-ray Magntic Circular Dichroism, first-principles calculations}
\maketitle
\section*{Introduction}
A magnetic tunnel junction becomes a multiferroic one if the
insulating (oxide) barrier is replaced by a ferroelectric. In addition
to the tunneling magnetoresistance (TMR), which is detected by
switching the electrode magnetizations from parallel
(${\uparrow\uparrow}$) to antiparallel (${\downarrow\uparrow}$), the
multiferroic tunnel junction (MFTJ) exhibits also tunneling
electro-resistance upon electric polarization reversal. In
consequence, the resistance ($R$) takes four different values, paving
the way for nonvolatile four-state memory
devices\cite{Velev2009,Garcia2010}. One prototype example is the
Co/PZT/LSMO system where abbreviation PZT is used for the ferrolectric
PbZr$_{0.2}$Ti$_{0.8}$O$_{3}$ oxide and LSMO for the ferromagnetic,
almost half metallic, La$_{2/3}$Sr$_{1/3}$MnO$_3$
oxide~\cite{Pantel2012}.  For a 3.2-nm-thick PZT barrier, epitaxially
grown on a thin LSMO film, it was shown that when the PZT polarization
changes from pointing toward LSMO to pointing toward Co, the TMR which
is given by TMR=
$(R_{\downarrow\uparrow}-R_{\uparrow\uparrow})/R_{\downarrow\uparrow}$
changes sign from $\unit+4{\%}$ to $\unit-3{\%}$. Until now, the
origin of this TMR sign inversion is not completely understood. Pantel
et al.~\cite{Pantel2012} argued that the inversion of the TMR sign
could be related to the inversion of the spin polarization at the
LSMO/PZT interface or, alternatively, at the Co/PZT
interface~\cite{Belashchenko05,Heiliger05}, while resonant
tunneling~\cite{Yuasa02,Wunnicke02b,Tsymbal2003} via the barrier was
ruled out. With regard to the Co/PZT interface, \textit{ab initio}
calculations \cite{Borisov2014} assumed that its geometric structure
is characterized by the terminating Ti(Zr)O$_{2}$ plane with
interfacial Co atoms residing on top of the oxygen atoms.  This
interface structure model has been frequently used in calculations,
but a clear-cut experimental proof has not been given up to now.
Several previous studies on different metal/oxide interfaces have
indicated that metals deposited on oxides have a strong tendency to
form oxide-like interfaces such as e.g. in the case of
Fe/MgO~\cite{Meyerheim2001, Tusche2005} and of
Fe/BaTiO$_{3}$~\cite{Bocher2012}, the latter being similar to the
Co/PZT system. For iron grown on the perovskite (PV)-type
ferroelectric BaTiO$_{3}$ (BTO) transmission electron microscopy
\cite{Bocher2012} has suggested the formation of a single FeO layer at
the interface, but very recently, a first-principles study by Imam et
al.\cite{Imam2019} of the Co/PZT interface proposed a more complex
interface defect structure characterized by an oxygen deficient
TiO$_{2}$ termination layer which is followed by an interstitial
oxygen atom residing on top of titanium. Depending on the polarization
state of the ferroelectric the vertical distance between the Ti-O
complex and the first cobalt layer strongly varies, which is suggested
to be responsible for the spin-polarization to change sign upon
polarization reversal~\cite{Imam2019}. Up to now, no direct and
independent proof for any of the proposed interface models has been
given. This calls for the application of well-established structure
analytical tools which are sensitive to monolayers of materials and
which are capable to study both long and short range order.

To this end we have carried out surface x-ray diffraction (SXRD) and
extended x-ray absorption fine structure (EXAFS) experiments on the
Co/PZT/LSMO system. These two techniques are complimentary to each
other with regard to the structural order probed, long range for SXRD
and short range in the case of EXAFS. In addition, the magnetic
properties were studied by x-ray magnetic circular dichroism (XMCD)
above the cobalt and titanium L-$_{II,III}$ edges in combination with
state of the art ab-initio calculations using the interface structure
model as input.  The magnetic moments of the Co and Ti ions at the top
interface were derived.

\section*{Results}

\subsection{Surface X-Ray Diffraction}

\begin{figure*}[t]
\centering \includegraphics[width=.9\textwidth]{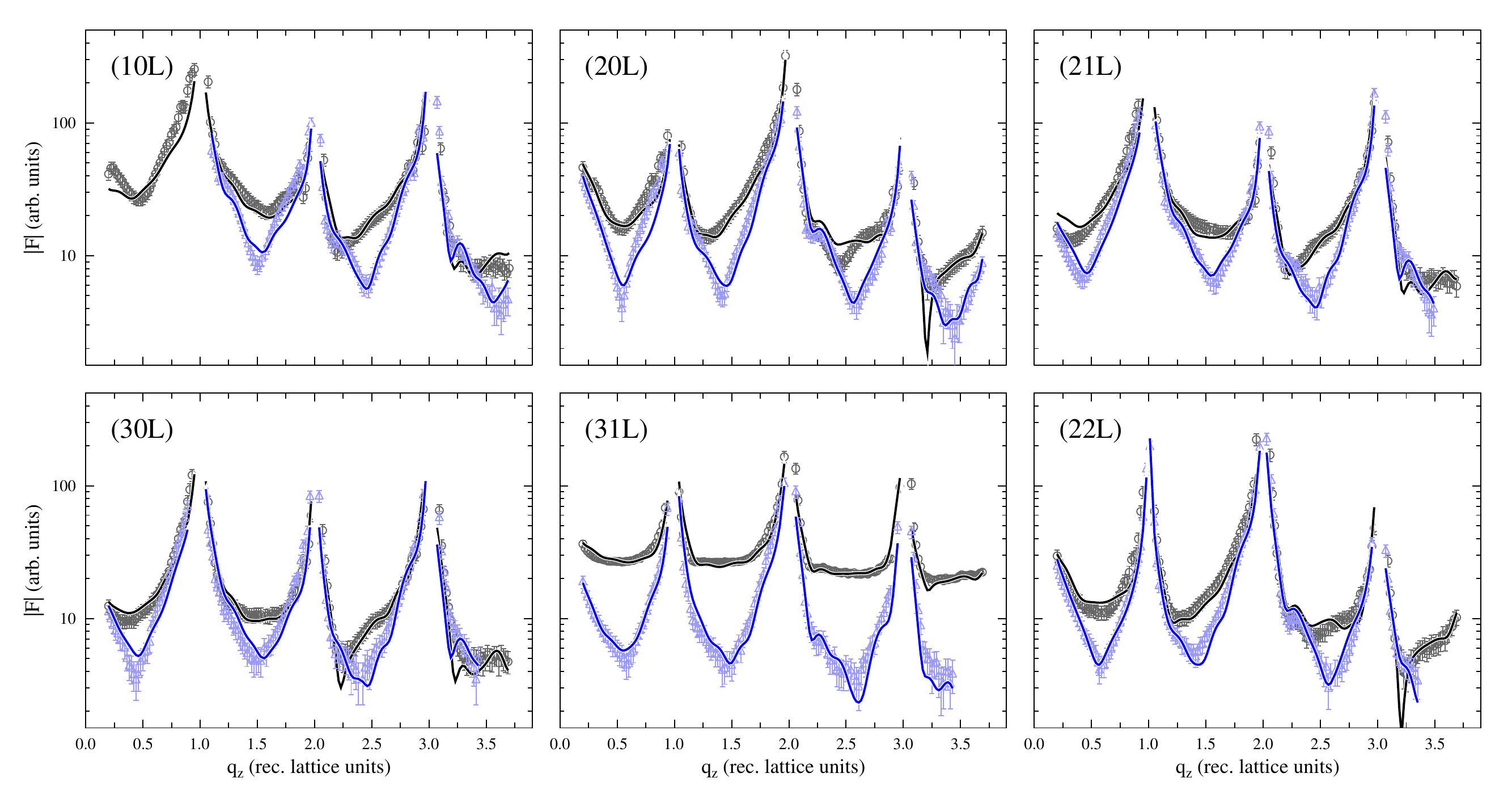}
\caption{\small (a-f): Experimental (symbols) and calculated (lines)
  structure factor magnitudes ($\mid$ F$_{obs}$$\mid$ ) along six
  symmetry independent crystal truncation rods collected for the
  PZT/LSMO sample (blue) and after in-situ deposition of approximately
  1~monolayer of cobalt (black). Bulk Bragg-reflections are located at
  integer $q{_z}$. Standard deviations ($\sigma$) of the $\mid$
  F$_{obs}$(HKL)$\mid$ are represented by the error bars.}
\label{rodsnew}
\end{figure*}

The LSMO (20~nm) and PZT (2 unit cells) films were grown on a single
crystalline SrTiO$_{3}$(001) [STO(001)] substrate by pulsed laser
deposition as discussed earlier~\cite{Pantel2012}. The SXRD
experiments were carried out at the beamline ID03 of the European
Synchrotron Radiation Facility (ESRF) in Grenoble (France) using a
ultra-high-vacuum (UHV) diffractometer and a two-dimensional (2D) pixel
detector. After transfer into the UHV system the sample was heated in
oxygen atmosphere (p$_{O2}$ $\approx$ 10$^{-6}$ mbar) up to about
300$^{\circ}$ C to remove carbon contamination and water. The
integrated intensities were recorded under grazing incidence
($\alpha$$_{i}$=1$^{\circ}$) of the incoming x-ray beam
($\lambda$=0.50~\AA).

Data were first collected for pristine PZT/LSMO/STO prior to cobalt
deposition. In the next step, cobalt was deposited on the pristine
sample using an electron beam evaporator. We have investigated the
structure of the interface and the films after depositing about 0.5
and 1.0 monolayers (ML) of cobalt. Here and in the following we refer
to a coverage of 1~ML as one adatom per surface unit cell, i.e.
6.55$\times$10$^{14}$ atoms/cm$^{2}$.  The cobalt film thickness was
calibrated by monitoring the intensity of the (1 0 1.5) anti-phase
reflection of the crystal truncation rod (CTR) versus deposition time
and by Auger electron spectroscopy (AES).  The amount of
\textit{long-range ordered} cobalt on the PZT surface is derived $a$
$posteriori$ by the development of the structure model based on the
least squares fit of the structure factor magnitudes as will be shown
below. In the following we discuss the analysis and the structure of
the sample covered by 1~ML of cobalt. The structure of the sample
studied after depositing half a ML is outlined in the Supporting
Information.

\begin{figure*}[ht]
\centering \includegraphics[width=.9\textwidth]{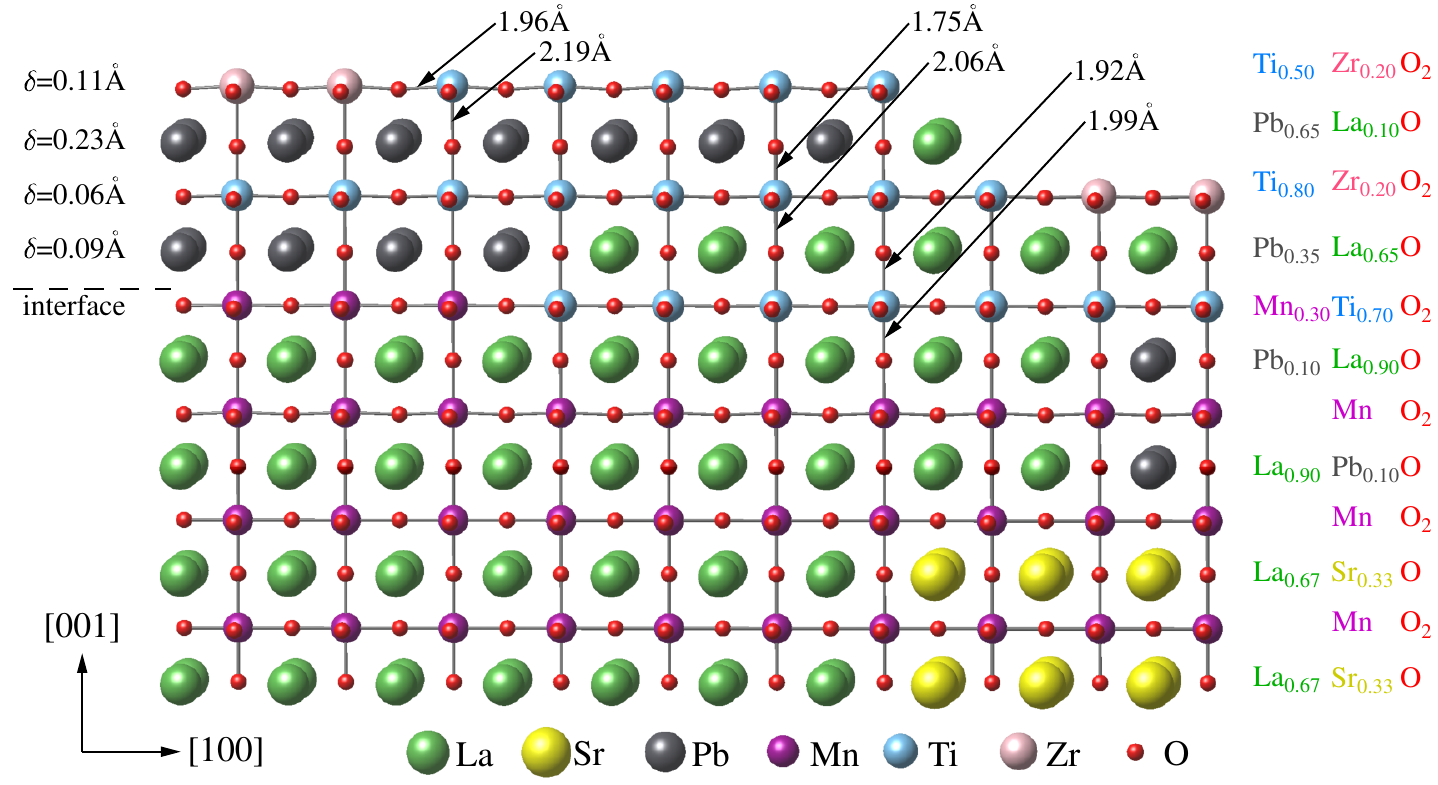}
\caption{Model of the pristine PZT/LSMO interface derived from
  the least squares fit of the CTR data shown in Figure 1 (blue
  curves). Colored balls represent atoms as indicated in the legend.
  Numbers indicate distances in \AA ngstr{\o}m units. The approximate
  stoichiometry in each layer is given on the right and is represented
  by the number of balls. }
\label{structure_pristine}
\end{figure*}

In Figure~\ref{rodsnew} several symmetry independent CTRs for the
pristine PZT/LSMO/STO (blue) and the cobalt covered (black) sample are
shown. Symbols represent experimental structure factor magnitudes
($\mid$ F$_{obs}$(HKL)$\mid$). Data collection was carried out by
performing line scans in reciprocal space along the q$_{z}$ direction
using a proper region of interest for the signal and background in the
2D pixel detector. For more details we refer to Ref.~\cite{Drnec2014}.
Lines in Figure~\ref{rodsnew} correspond to calculated structure
factor magnitudes ($\mid$ F$_{cal}$(HKL)$\mid$ ). Each data set
consists of about 2000 reflections which are averaged to about 1000 by
symmetry equivalence based on the $p4mm$ plane group. The standard
deviation $\sigma$ of the $\mid$ F$_{obs}$(HKL)$\mid$, estimated from
symmetry equivalent reflections, is about 10\% which is represented by
the error bars in Figure~\ref{rodsnew}. The coherent addition of the
adlayer structure factor to the bulk CTR structure factor allows the
precise analysis of the adlayer atomic arrangement even in the case of
an absence of superstructure.~\cite{Meyerheim2018}

The least squares fit of $\mid$ F$_{cal}$(HKL)$\mid$ to $\mid$
F$_{obs}$(HKL)$\mid$ was carried out by using the program
"Prometheus"~\cite{Zucker1983}. Owing to the high symmetry of the
structure (plane group $p4mm$) only the z-parameters of the atoms need
to be refined while the x- and y coordinates are fixed by symmetry.
For instance, the La(Pb) and Mn(Ti) atoms reside in Wyckoff positions
1a and 1b, respectively, while the two sets of oxygen atoms are
located in 1b and 2c. In the analysis we have considered two unit
cells (uc) of PZT in addition to two uc of the LSMO substrate. Deeper
layers in the LSMO film were treated as bulk like. In addition to the
positional parameters an overall scale factor and Debye-parameter (B)
was refined, the latter representing static and dynamic
disorder~\cite{Kuhs1992}.

At first we discuss the pristine PZT/LSMO sample.

\begin{figure*}[ht]
\centering \includegraphics[width=0.9\textwidth]{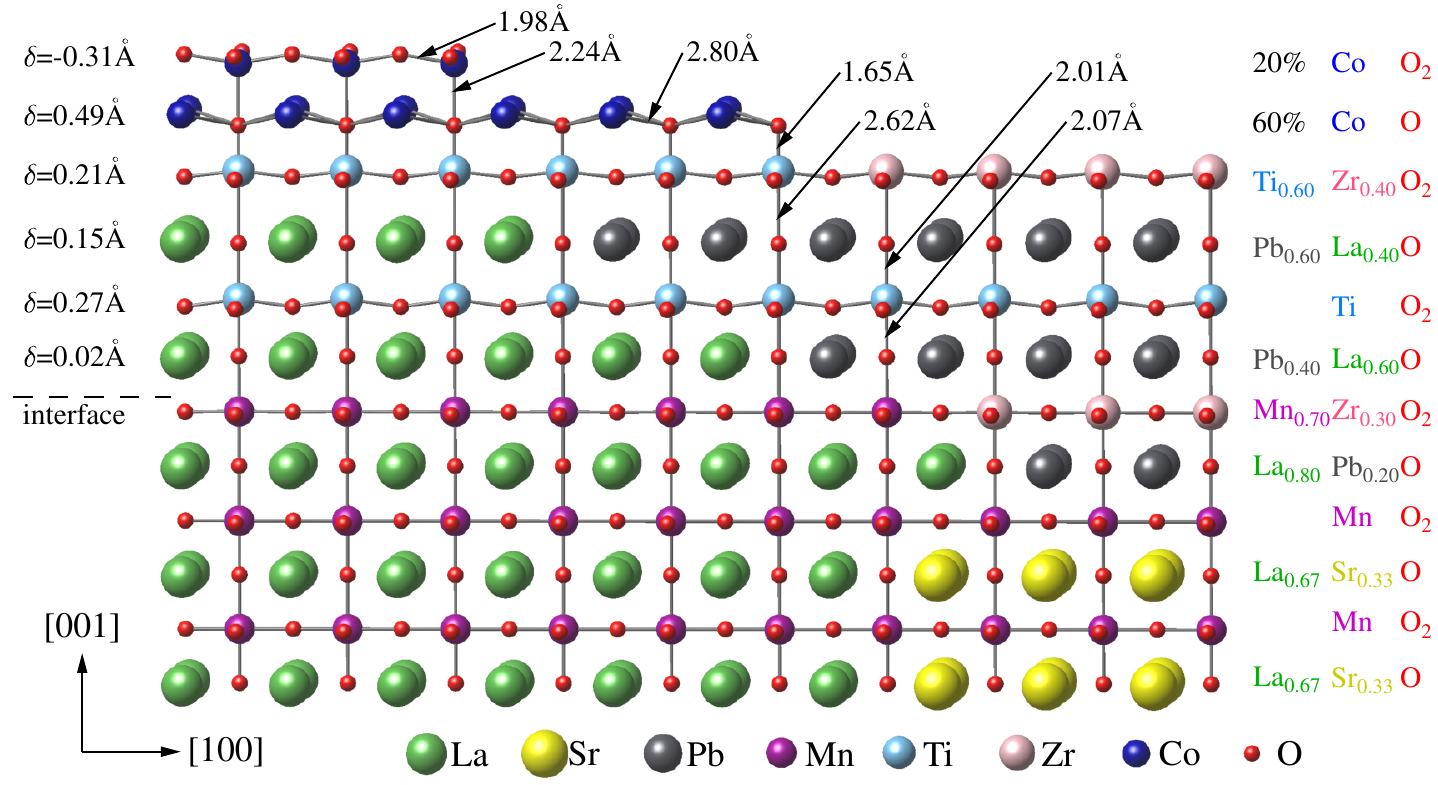}
\caption{Model of the Co/PZT/LSMO interface derived from the least
  squares fit of the CTR data shown in Figure 1 (black curves).
  Colored balls represent atoms as indicated in the legend. Numbers
  indicate distances in \AA ngstr{\o}m units. The approximate
  stoichiometry in each layer is given on the right and is represented
  by the number of balls.}
\label{structure_cobalt}
\end{figure*}

Fig.~\ref{structure_pristine} shows the structure model viewed onto
the (010) plane. Atoms are represented by colored balls as labelled in
the legend to the figure. For better visualization of the partial
occupancies the number of balls approximately represents the
fractional coverage within each layer. We find that the PZT film is
terminated by a Ti(Zr)O$_{2}$ layer, which is incomplete (fractional
occupancy about 70\%). Secondly, we find a considerable degree of
cationic intermixing at the PZT/LSMO interface which is highlighted by
the dashed line. Across the interface, cationic exchange of lead (Pb)
versus lanthanum (La) and titanium (Ti) versus manganese (Mn) is
observed.  In this respect the PZT/LSMO interface resembles the
LaFeO$_{3}$/SrTiO$_{3}$ interface investigated
previously~\cite{Xu2017}. The approximate composition of the
individual layers is given on the right. The cationic exchange appears
as not to be symmetric at least as far as the Pb/La exchange is
concerned for which the uncertainty is in the $\pm$0.2 range, while
for the Mn/Ti exchange it is larger in the $\pm$0.4 range due to the
similar scattering amplitudes of Ti and Mn.

Metal-oxygen distances are typical for those found in PV-type
structures. For the octahedrally coordinated metal atoms (Ti, Zr) we
derive values around the 2~\AA~ in general with two exceptions located
in the terminating range where some vertical relaxation is observed
leading to a deviation of approximately 0.10-0.15~\AA~ from the
"ideal" value of 1.95~\AA~if we refer to a perfect cubic STO lattice
parameter a$_{0}$=3.905~\AA. Similarly, for the 12-fold coordinated
atoms (La, Pb), distances in the 2.7~\AA~range are derived (not
indicated in detail in Figure~\ref{structure_pristine}). The accuracy
of the distance determination is estimated to lie in the $\pm$
0.05~\AA~range. The ferroelectric polarization is related to a
intra-layer vertical shift ($\delta$) between the cation and oxygen
which can be positive ($\delta >0$ for P$\uparrow$) or negative
($\delta >0$ for P$\downarrow$). For the 2 uc thick PZT film we find
positive values for the as-deposited state, which increase from 0.09
and 0.06~\AA~in the first u.c. at the interface to LSMO to 0.23 and
0.11~\AA~in the top u.c. for the PbO and the Ti(Zr)O$_{2}$ layer,
respectively (see Figure~\ref{structure_pristine}). These magnitudes
are smaller than those theoretically predicted
($\approx$0.3~\AA)~\cite{Borisov2014}, which -however- refer to the
bulk structure. With this model an excellent fit (solid blue lines) to
the $\mid$ F$_{obs}$(HKL)$\mid$ is achieved characterized by the
un-weighted residuum R$_{U}$=0.13 and a goodness of fit
GOF=1.67\cite{Ru}.

In the next step, cobalt was deposited on the PZT surface while
simultaneously monitoring the (1 0 1.5) reflection intensity at the
anti-phase scattering condition along the CTR. The Co deposition was
stopped at intensity saturation corresponding to an 300\% increase
relative to the pristine sample. In general, there is a significant
overall increase of the $\mid$ F$_{obs}$(HKL)$\mid$ along the CTRs.
Based on AES, the cobalt coverage is equal to approximately 1~ML. At
this film thickness some fraction of the deposited cobalt might not be
adsorbed in a long range ordered structure. In order to fit the
experimental data several models including those suggested by
different authors~\cite{Pantel2012,Bocher2012,Imam2019} were
considered as starting models followed by relaxation of the
z-positions and variation of the site occupancies. The best fit, which
is given by the black solid lines in Figure~\ref{rodsnew}, is achieved
for the model which is outlined in Figure~\ref{structure_cobalt}.

The structure is characterized by two fractionally filled layers of
cobalt oxide which continue the PV-type structure of the substrate. In
detail, a 60\% filled atomic layer of CoO follows on the Ti(Zr)O$_{2}$
terminating layer of the PZT film. Above the CoO layer a 20\% filled
atomic layer of CoO$_{2}$ is located. In both cobalt oxide layers a
relatively large $\delta$ parameter is found (0.49 and -0.31~\AA~for
CoO and CoO$_{2}$, respectively), while interatomic Co-O distances are
typical for those in PV-type structures. There is one quite short
(1.65~\AA) oxygen-metal distance between the CoO-layer oxygen and the
topmost Ti(Zr) ions, which is also fairly well reproduced for the 0.5
ML Co/PZT interface, which is discussed in more detail in the
Supplementary Information. The $\delta$ parameters within the
Ti(Zr)O$_{2}$ layers of the PZT film are enhanced relative to the
pristine one (e.g. 0.27 versus 0.06~\AA~ and 0.21 versus 0.11~\AA).
The structure of deeper layers involving PZT and LSMO are not
significantly modified as compared to the PZT/LSMO sample.

The fit to the $\mid$ F$_{obs}$(HKL)$\mid$ is represented by the black
solid line and is slightly worse than that for the pristine one
(R$_{U}$=0.22, GOF=1.71). This is almost completely due to the bad fit
at low values of q$_{z}$ along the (10L) CTR as a result of problems
with the background subtraction (see also Supplementary information). 
We nevertheless included these data
into the analysis. Here we recall that the comparison between the CTRs
of the pristine and the cobalt dosed sample in Figure~\ref{rodsnew}
impressively points out that the CTRs profiles extremely sensitively
depend on the variation of the structure leaving no room for different
film and interface models.

In summary, we conclude that adsorption of a approximately 1~ML of
cobalt on the PZT/LSMO surface leads to the formation of a PV-like
cobalt oxide (CoO-CoO$_{2}$ sequence) forming islands on the PZT
surface which cover approximately 60\% (first layer) and 20\% (second
layer) of the surface. The SXRD analysis finds no evidence for the
presence of cobalt metal at this coverage. This conclusions are
supported by the structure analysis of a Co/PZT/LSMO sample with a
cobalt film thickness of approximately 0.5~ML, i.e. about one half of
the coverage discussed here (see Supporting Information). For this
sample we find the formation of only the first CoO-type of layer with
a fractional coverage of about 50\%.

\subsection{Extended X-Ray Absorption Fine Structure (EXAFS)}

The EXAFS measurements were carried out in-situ at the Samba beamline
of the Synchrotron Radiation Facility Soleil in Saint Aubin (France)
using an end station equipped with standard surface analytical tools.
After cobalt deposition on the PZT/LSMO sample, EXAFS data were
collected above the Co-K absorption edge (E$_{0}$=7709~eV) in the
fluorescent yield (FY) mode using a single element silicon drift
detector (Bruker) placed in plane at an angle of 90$^{\circ}$ to the
incoming beam. The in-plane and out-of-plane cobalt atomic environment
was investigated by exploiting the polarization dependence of the
EXAFS signal, taking advantage of the linear polarization of the x-ray
beam. EXAFS spectra were collected with the electric field vector
aligned either perpendicular (E$_{\perp}$) or parallel
(E$_{\parallel}$) to the plane of the substrate \cite{Koningsberger}.

\begin{figure*}[ht] \centering
  \includegraphics[width=.9\textwidth]{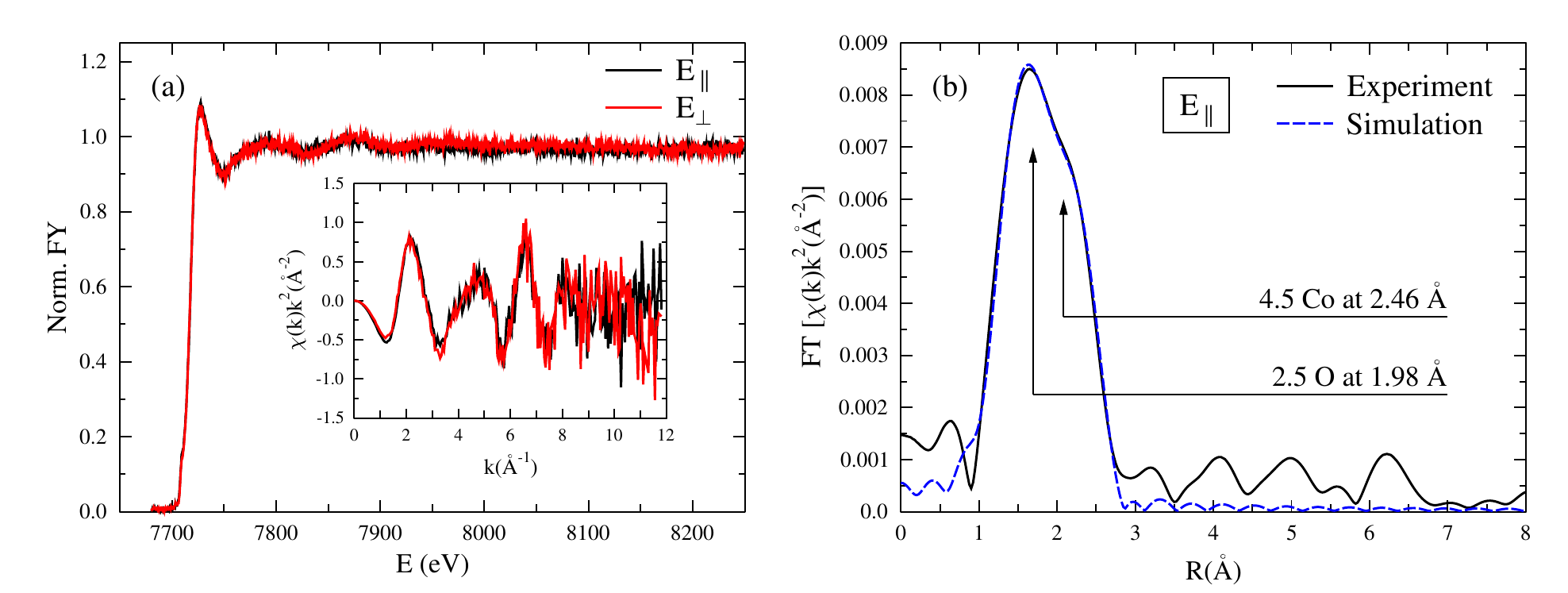} \caption{ (a): Normalized
    fluorescence yield collected for about 1 ML of cobalt deposited on
    PZT with the electric field vector parallel (E$_{\parallel}$) and
    perpendicular (E$_{\perp}$) to the sample surface. The inset shows
    the k$^{2}$-weighted interference function [$\chi(k)\times
    k^{2}$]. The absence of any polarization dependence indicates a
    nearly isotropic environment. (b): Fourier-Transform of the
    k$^{2}$-weighted interference function [$\chi(k)$] for the
    E$_{\parallel}$ geometry.  The experimental and fitted curve is
    represented by the full black and dashed blue line, respectively.}
\label{exafs1}
\end{figure*}

The sample was prepared in the same way as done for the SXRD
experiments and about 1 ML of cobalt was deposited in-situ.
Figure~\ref{exafs1}(a) shows the normalized FY, which is proportional
to the linear absorption coefficient $\mu$, versus photon energy for
both geometries. Direct inspection reveals that both curves are nearly
identical indicating an isotropic environment around the cobalt atoms.
This is also evident by inspecting the k$^{2}$-weighted interference
function [$\chi(k)$] (see inset of Figure~\ref{exafs1}(a), which is
obtained after background subtraction and multiplication with k$^{2}$,
the squared wave vector magnitude of the photoelectron. The magnitude
is given by: $k=\sqrt{(2m/\hbar) (E-E_{0})}$ with E and E$_{0}$ being
the photon and the absorption edge energy, respectively. The
quantitative analysis begins with the evaluation of the magnitude of
the Fourier-Transform (FT) of the $\chi(k)\times k^{2}$, which is
shown in Figure~\ref{exafs1}(b) for E$_{\parallel}$ geometry.

The dominant peak in the FT in the 2 \AA ngstr{\o}m range has an
asymmetric shape suggesting that it contains the contribution of two
atomic shells. This peak was fitted in R-space using the program
Feff~\cite{FEFF}, based on theoretical scattering amplitudes and
phases. We find that the first shell corresponds to approximately 2.5
oxygen atoms at a distance of 1.98~\AA~, while the second shell is
composed of 4.5 cobalt atoms as a distance of 2.46~\AA. The fit
parameters are listed in Table 1.

\begin{table}
  \caption{Table of structural parameters for approximately 1~ML of cobalt on
    PZT/LSMO. The meaning of the parameters is as follows: R:=
    refined neighbor distance, N$^{*}$:=effective polarization
    dependent coordination number for S-edge EXAFS, $\sigma^{2}$:=mean
    squared relative displacement amplitude, $\Delta$E$_{0}$:=shift of
    absorption edge (identical for both shells), R$_{u}$:=Residual in percent~\cite{RUEXAFS}. The
    amplitude reduction factor (S$_{0}^{2}$) was kept constant at
    S$_{0}^{2}$=0.90 in all cases. Parameters labelled by an asterisk
    (*) are kept fixed. Uncertainties are given in brackets.}
   \begin{center}
   \begin{tabular}{l|c|c|c|c|c|c}

     Pol. & Shell & R (\AA) & N$^{*}$ & $\sigma^{2}$(\AA$^{2}$)&$\Delta$ E$_{0}$               (eV)& R$_{u}$ \\[0.5ex]
     \hline\hline
    \toprule
    E$\perp$  & O1 & 1.94(3)  & 2.0(5)   & 0.017(5) & -1.2   &  4.8  \\
              & Co & 2.43(3)  & 4.5(5) & 0.013(5)   & -1.2   &       \\
    \hline\hline
    E$\parallel$  & O1 & 1.98(3) & 2.5(5)    & 0.012 (5) & 3.1 &  4.0 \\
                  & Co & 2.46(3) & 4.5(5)    & 0.017 (5) & 3.1 &
                \end{tabular}
                \end{center}
\end{table}

This model is similar to that derived by SXRD, indicating the formation
of cobalt oxide. The difference is that the EXAFS analysis in addition
finds a cobalt metal-like shell. Compared to the bulk hcp-Co nearest
neighbor distance of 2.51~\AA, the distance of 2.46(3)~\AA~ between Co
atoms corresponds to a contraction which is related to the "mesoscopic
misfit" owing to the reduced
coordination~\cite{Mironets2008,Meyerheim2012,Brovko2014}. Also, the
EXAFS analysis provides no clear evidence for the presence of the
second Co-O shell that should be expected in the distance range
between 2.60 and 2.80~\AA~ characteristic for the cation in the
12-fold coordination of the PV structure. We believe that this is due
to the small contribution of this shell in comparison with the strong
Co-Co metal contribution at 2.46~\AA.

We estimate the uncertainties for the effective coordination numbers
(N$^{\ast}$) to lie in the 10 to 25\% range, those of the distance
determination to be 0.03~\AA~ at most. The amplitude reduction factor
(S$_{0}^{2}$=0.90), which is important for the correct determination of
N$^{\ast}$ was derived by using a CoO standard (sodium chloride-type
structure) prepared by annealing the as deposited film in oxygen
atmosphere (see Supporting Information).

\subsection{X-Ray Circular Magnetic Dichroism}

The XMCD experiments were carried out at the Boreas bemaline at the
Synchrotron Radiation Facility Alba (Barcelona, Spain) using the
HECTOR vector magnet end station. Cobalt was deposited on the PZT/LSMO
sample in approximately the same film thickness regime as in the SXRD
and EXAFS experiments ($\approx$ 1~ML). The absorption coefficient in
the vicinity of the Co-L$_{II,III}$ edge (E$_{0}$=793 and 778~eV,
respectively) was monitored via the total electron yield method at
approximately 4~K by collecting the sample drain current.

\begin{figure*}[ht]
  \centering \includegraphics[width=.9\textwidth]{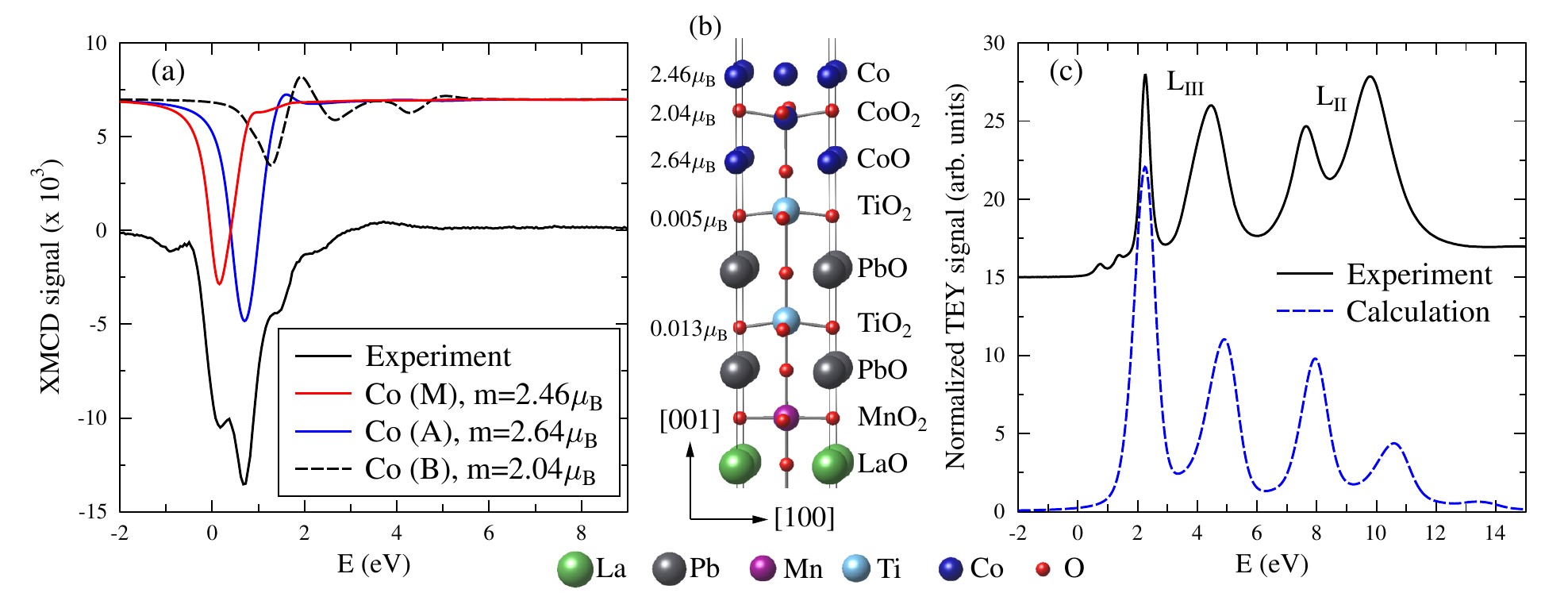}
  \caption{(a): Experimental (lowest curve) and calculated XMCD signal
    in the vicinity of the the Co-L$_{II,III}$ edge (shifted for
    clarity). In total three different contributions to the line shape
    are found which are attributed to (i) CoO at the interface to the
    TiO$_{2}$ layer (A), to (ii) CoO$_{2}$ in the second layer (B) and
    to metallic cobalt (M). The structure model is shown schematically
    in (b). (c): Experimental (symbols) and calculated XAS signal at
    the Ti-L $_{II,III}$ edge. Titanium atoms in the first and second
    layer from the interface acquire a magnetic moment of 0.005 and
    +0.013 $\mu_{B}$, respectively.}\label{xmcd1}
\end{figure*}

Figure ~\ref{xmcd1}(a) shows the experimental XMCD spectrum obtained
for 1 ML Co/PZT/LSMO red in a magnetic field of 6T. At the
L$_{III}$ edge the lineshape is characterized by three components
which appear within about 2 eV and which are clearly separable.
Qualitatively these can be attributed to the different cobalt
environments at the Co/PZT interface in agreement with the SXRD and
EXAFS analysis. In order to quantitatively investigate the XMCD
spectrum, we employed first-principles simulations applying a fully
relativistic linear muffin-tin orbital (LMTO)
method~\cite{Andersen1975, PYLMTO, NKA+83,MPK80}.  The implementation
of these methods is based on four-component basis functions
constructed by solving the Dirac equation inside a Wigner-Setz
cell\cite{NKA+83}. This is decisive for a correct description of
$p_{1/2}$ states of heavy elements such as Pb or Bi \cite{MPK80}. The
X-ray absorption and dichroism spectra were calculated taking into
account the core level exchange splitting. The finite lifetime of a
core hole was accounted for by folding the spectra with a Lorentzian
based on core level widths $\Gamma_{L_{2,3}}$ taken from
Ref.~\onlinecite{CaPa01}. Spectrometer resolution was included by a
Gaussian function of width 0.6 eV. The positions of main peaks of
calculated spectra (shifted for clarity) are in a good agreement with
experimental results, although the width of the intensities cannot be
reproduced correctly because of a one-particle approximation used in
our simulations.

XMCD spectra were calculated taking into account the components
derived from the SXRD and the EXAFS analysis and compared with the
experimental spectrum as shown in Figure~\ref{xmcd1}(a). The somewhat
simplified structure model is shown in Figure~\ref{xmcd1}(b).  In
addition, the X-ray absorption (XAS) at the Ti-L$_{II,III}$ edge
(E$_{0}$$\approx$460 and 454~eV, respectively) is shown in
Figure~\ref{xmcd1}(b). Direct inspection of the calculated XAS \& XMCD
spectra indicates that the different components fit to the presence
of cobalt metal (M), cubic-like CoO (A) and CoO$_{2}$ (B). The local
magnetic moment of cobalt is found to be equal to 2.46 $\mu_{B}$
whereas the cobalt ions bear a magnetic moment of 2.64 $\mu_{B}$ in
the CoO layer and 2.04 $\mu_{B}$ in the CoO$_{2}$ layer.  The magnetic
moments (m) of the Ti ions within the PZT film are found
to be equal to only m=+0.005 (first layer) and to m=+0.013 $\mu_{B}$ (second layer).

\section*{Discussion and Conclusion}

We have presented a thorough structural and spectroscopic
investigation of the multiferroic Co/PZT/LSMO system, analyzing the
two interfaces and in particular the evolution of the top interface
upon deposition of cobalt. Using complementary techniques (SXRD and
EXAFS) clear evidence was given that the two interface structures are
rather complex. Importantly, the top interface consists of different
components of the cobalt environment. About half a monolayer of cobalt
atoms gets oxidized forming a PV-like continuation of the crystal
structure. In detail, a cubic-like CoO layer follows the TiO$_{2}$
layer terminating the PZT film followed by a CoO$_{2}$ layer.
Additionally deposited cobalt leads to the rapid evolution of a
metallic cobalt film, as probed by EXAFS, which is sensitive to the
short range order, but which escapes the SXRD analysis. XMCD and XAS
support these results indicating that the local magnetic moments of
cobalt within the cobalt metal/cobalt-oxide interface layer lie in the
range between 2.0 and 2.5 $\mu_{B}$.

Our calculations from first principles suggest that the Co$_2$O$_3$ PV
layer represents a metallic, ferromagnetic and tetragonally compressed
phase, the latter because of the lattice mismatch with the PZT/LSMO
substrate. In detail, the PV-type cobalt metal/cobalt-oxide overlayer
induces a magnetic moment of +0.005 $\mu_{B}$ on the interfacial Ti
atoms in the 2 uc thick PZT barrier.  The second layer Ti atoms
exhibit a slightly larger magnetic moment of +0.013 $\mu_{B}$, which
-however- is induced by the LSMO film underneath. The magnitude of
this moment can be varied by the PZT polarization
direction~\cite{Borisov2017} which requires a minimum PZT barrier
thickness of 4 uc. In summary, we find from the analysis of the XAS
data, that the titanium moments are parallel to those of the cobalt
atoms within the cobalt/cobalt-oxide layers, albeit they are very
small in general. This is at variance with the magnetic properties
predicted for several metal/PV interface models discussed
previously~\cite{Meyerheim2011,Valencia2011,Bocher2012,Meyerheim2013,Borisov2014,Borisov2017,Imam2017,Imam2019},
where an anti-ferromagnetic alignment of the titanium moments has been
found. The main reason for this effect is the formation of strong
in-plane Co-O bonds within the CoO and the CoO$_2$ layer, which owing
to the increased localization of the Co $d_{xz}$, $d_{yz}$ and
$d_{z^2}$ orbitals involved reduce the overlap of the corresponding
charge density with the titanium orbitals.

We conclude that the complex Co/PZT interface involving a single unit
cell of a PV-type cobalt-oxide plays a key role for the
electronic and transport properties of this multiferroic tunnelling
junction. Apart from the ferromagnetically aligned interfacial Ti
atoms, there is another important factor influencing its functionality
which originates by the polarization-dependent partial metallization
of the PZT film near its interfaces.~\cite{Quindeau2015} In this
context, the presence of the conducting and strongly ferromagnetic
PV-like cobalt-oxide layer can crucially affect the measured TMR of
Co/PZT/LSMO.  Thus, the complex interface involving an oxide interface
is expected to have important consequences on the electronic, magnetic
and the transport properties.

\section*{Methods}

\subsubsection*{Sample preparation}

The oxide films were epitaxially grown on on 0.1$^{\circ}$ miscut
SrTiO$_{3}$(001) substrates by pulsed laser deposition (PLD) using
stoichiometric ceramic targets under oxygen atmosphere
(p$_{O_{2}}$=0.2 mbar) while substrates were kept at 550-600$^{\circ}$
C. For the PLD process an excimer laser ($\lambda$=248 nm) with a
fluence of 300 mJ/cm$^{2}$ at the sample position was used. Cobalt
deposition was carried out in-situ prior to the SXRD, EXAFS and XMCD
experiments by evaporation from a cobalt rod heated by electron
bombardment.

\subsubsection*{Surface X-ray diffraction}

Surface X-ray diffraction experiments were carried out at the beamline
ID03 of the European Synchrotron Radiation Facility in Grenoble
(France) using a six-circle UHV diffractometer using a primary beam of
24~keV ($\lambda$=0.5~\AA) photon energy and grazing incidence
($\alpha_{i}$=1$^{\circ}$ of the beam to the sample plane. The
intensity distribution [I$_{obs}$(HKL)] along the q$_{z}$=L$\times$ c*
direction in reciprocal space was monitored by line scans using a
two-dimensional (2D) pixel detector as discussed in
Refs.~\cite{Schleputz2005,Drnec2014}. For each data set approximately
2000 reflections were collected along 12 CTRs which reduce to
approximately 1000 along six CTRs by symmetry equivalence (plane
symmetry group $p4mm$). The I$_{obs}$(HKL) were then multiplied by
instrumental correction factors C$_{corr}$ as outlined in detail in
Refs.~\cite{Vlieg1997,Drnec2014} yielding the experimental structure
intensities [$\mid$ F$_{obs}$ (HKL)$\mid$ $^{2}$] via: $\mid$
F$_{obs}$(HKL)$\mid$ $^{2}$= I$_{obs}$(HKL) $\times$ C$_{corr}$. The
total uncertainty (1$\sigma$) is estimated by the quadrature sum of
the statistical and the systematic error as outlined e.g. in
Ref.~\cite{Robinson1992}. We find the reproducibility of symmetry
equivalent $\mid$F$_{obs}$(HKL)$\mid$ $^{2}$ is equal to 10\%.

\subsubsection*{EXAFS}
EXAFS data have been recorded at the SAMBA beamline of SOLEIL
synchrotron (Saint Aubin, France). Monochromatization has been
performed using a Si (220) monochromator, two Pd coated mirrors
performed harmonic rejection (6mrad grazing incidence). Monochromatic
flux ($I_0$) has been measured with a ionisation chamber placed before
the SurfAS UHV setup. Co was evaporated from a 1 mm Co rod with an
e-beam evaporator (Omicron) and deposited thickness has been
calibrated with a quartz micro balance (Sycon STM-2XM) and checked by
Auger (CMA100, Omicron). Fluorescence yield was measured with a single
element silicon drift detector (Bruker) placed in plane at an angle of
90$^{\circ}$ to the incoming beam.

\subsubsection*{XMCD}
\subsubsection*{Experiments}

The XAS and XMCD measurements were carried out at BOREAS beamline at
the ALBA synchrotron light-source in the high-field magnet endstation
HECTOR using fully circularly-polarized photons produced by an
APPLE-II type undulator.~\cite{barla2016} The total electron yield
(TEY) signal was measured as the sample-to-ground drain current and
normalized by the incident photon flux determined by the TEY signal
measured on a freshly-evaporated gold mesh placed between the last
optical element and the sample. Both TEY signals were amplified by a
Keythley model 428 electrometer. During the XMCD experiments a
magnetic field of 6T collinear with the photon incidence direction was
applied, generated by a superconducting vector-electromagnet
(Scientific Magnetics). The XMCD spectra were obtained by measuring
the XAS signal with opposite elicities photons and calculating the
difference between them. The Co evaporation was carried out in the
magnet chamber keeping the sample at room temperature by an e-beam
evaporator (SPECS) using an high-purity Co-rod (Goodfellow). The
pressure in the measurement chamber was lower than
2$\times$10$^{-10}$mbar during the measurements and Co evaporation.
The Co coverage was estimated via the XAS edge-jump intensities using
previously calibrated values.

\subsubsection*{First-principles calculations}
Electronic structure calculations and simulations of XAS \& XMCD
spectra were performed using a fully relativistic LMTO
method~\cite{Andersen1975,PYLMTO,MPK80}, which is based on a
four-component basis functions constructed by solving the Dirac
equation inside a Wigner-Setz cell or an atomic sphere\cite{NKA+83}.
The crystalline structure obtained from the current experiments served
as input for our calculations, performed within the density functional
theory utilizing a generalized gradient approximation in the version
of Perdew, Burke, and Ernzerhof.~\cite{Perdew1996} The X-ray
absorption and dichroism spectra were calculated taking into account
the core level exchange splitting. The finite lifetime of a core hole
was accounted for by folding the spectra with a Lorentzian based on
core level widths $\Gamma_{L_{2,3}}$ taken from
Ref.~\onlinecite{CaPa01}.  The $k$-space integrations were performed
with the improved tetrahedron method.~\cite{Bloechl1994} To attain
good convergence of the total energy and of the simulated spectra, we
used 16$\times$16$\times$ divisions of the Brillouin zone.

\bibliographystyle{unsrt}
\bibliography{./CPZT}

\begin{thebibliography}{10}
\bibitem{RUEXAFS}
The unweighted residuum (Ru) is defined by:
  R$_u=\sum|(Yi_{obs}-Y_{calc})|/\sum(Y_{obs})$, where Y$_{obs}$ and Y$_{calc}$
  are the experimental and calculated magnitudes of the peak selected in the FT
  of the $\chi(k)$ spectrum and the summation extends over all points in the
  window used to filter the peak.
\end{thebibliography}
\section*{Acknowledgments}

HLM and KM thank the staff of the ESRF and ALBA for their hospitality
during their stay in Grenoble and Barcelona. AP and NJ acknowledge the
synchrotron SOLEIL for supplying beamtime.  Technical support by Frank
Wei{\ss} is gratefully acknowledged. This work is supported by the
Deutsche Forschungsgemeinschaft under grant SFB 762. The ALBA beamtime
access was made possible via 2015-IHR-MV and
official proposal number No. 2016021628.

\section*{Author contributions statement}

HLM and AP devised the X-ray diffraction, the EXAFS experiment, AQ
prepared the samples. HLM, KM  carried out the X-ray experiments at
the ESRF. Data analysis was done by KM and HLM. The EXAFS experiments
were carried out by AP, NJ and EF at Synchrotron Soleil. The XMCD
experiments at ALBA were performed by MV, HBV, PG and HLM. VA and AE
conducted the calculations to the XMCD spectra. HLM, AE, and MV wrote
the manuscript, with input from all authors. All authors have read and
approved the manuscript.
\section*{Additional information}
\textbf{Competing non-financial interests}
The authors declare no competing non-financial interests.
\end{document}